# Aperiodic lattices for tunable photonic bandgaps and localization

**Subhasish Chakraborty[*1], Michael C. Parker[2], Robert J. Mears[3]**

1. Microelectronics Research Centre, Cavendish Laboratory, Department of Physics, University of Cambridge, Madingley Road, Cambridge CB3 0HE, UK

2. Fujitsu Laboratories of Europe Ltd., Columba House, Adastral Park, Ipswich IP5 3RE, UK

3. Pembroke College, Trumpington Street, Cambridge CB2 1RF, UK

Photonic bandgap engineering using aperiodic lattices (AL's) is systematically studied. Up to now AL's have tended to be defined by specific formulae (e.g. Fibonacci, Cantor), and theories have neglected other useful AL's along with the vast majority of non-useful (random) AL's. Here we present a practical and efficient Fourier space-based general theory, to identify all those AL's having useful band properties, which are characterized by well-defined Fourier (i.e. lattice momentum) components. Direct real-space optimization of AL's tends to be computationally demanding, and is also difficult to generalise beyond 1D. However, via our Fourier space-based inverse optimization algorithm, we efficiently tailor the relative strength of the AL Fourier components for precise control of photonic band and localization properties.

---

[*]Present address: School of Electronic & Electrical Engineering, University of Leeds, Leeds LS2 9JT, UK, E-mail: s.chakraborty@leeds.ac.uk

In the design of photonic crystals used in photonic integrated circuits and dense wavelength division multiplexing (DWDM) systems, increasing importance is being placed on careful control of the transmission characteristics, which derive from the device band properties. There is thus a high premium on the ability to engineer the band properties of a photonic crystal in a systematic manner. Photonic crystals[1] are usually characterized by a well-defined lattice periodicity. However, photonic bandgap (PBG) engineering[2,3], for example to achieve field localization or fine tuning of the band properties, requires breaking of the lattice periodicity thorough the introduction of single or multiple defects (missing or extra scattering sites)[4-7]. Whilst this renders the lattice essentially aperiodic[8,9] in the most general sense, it is highly unlikely that a randomly chosen aperiodic lattice (AL) will have any useful field localization or band properties. Hence the question is how to identify the relatively small number of useful AL's (in terms of their band properties) from the very large number of possible AL's?

Conventional PBG engineering has tended to be based mostly on a 'forward' process of defining a lattice structure and then determining the band properties, e.g. through the use of constraining formulae such as the Fibonacci or Cantor based quasi-periodic lattices[10-12]; through coupled-cavity structures, proposed by Yariv *et al.*[6]; or by intuition (accumulated design experience) and trial-and-error, such as the high-Q cavity structures reported by Akahane *et al.*[13]. There has also been a growing interest in the 'inverse' process of determining appropriate aperiodic photonic lattices from the desired scattering properties (functionality)[14-18]. With the exception of quasiperiodic lattices, which can form stop-bands considerably away from the conventional Bragg frequency, these techniques modify the spectral functionality mostly within or around the Bragg stop-band. Also in most cases (both 'forward' and 'inverse') the underlying principle is the direct optimization of the band properties by characterization of the AL's in real-space. Even for a moderate number of scattering sites and defects these methods become computationally extremely demanding, and complex to generalise beyond 1D.

In this paper we believe we present the first general theory and systematic design tool to reveal a previously unknown landscape of essentially aperiodic lattices (AL's), which show useful and novel spectral (bandgap) functionalities, e.g. tuneable bandgaps, and multiple localized states. Previous theories have discarded these useful AL's along with the vast majority of non-useful (random) AL's, because they have characterized photonic lattices only with regular periodicity of the real space lattice. However, as we explain in our paper, we have adopted a Fourier-space ($k$-space) approach[19], where AL's having useful band properties are characterized by well-defined spatial frequency (i.e. lattice momentum) components. We have used a discrete Fourier transform (DFT)-based inverse optimization algorithm to tailor the Fourier components of an AL to match that of a "target" function; the target function, also of course, being defined in Fourier-space and equating to the desired band properties. Overall, the entire scheme turns out to be much more practical and efficient in achieving desired and controllable photonic band functionality. Indeed, we demonstrate that AL's[20-22] inherently offer the most flexible platform to achieve novel spectral functionalities. In this context, our method provides answers to questions such as: For a given overall length and lattice constant $\Lambda$ of a 1D lattice, what are the various defect combinations that produce discontinuities along the angular frequency $w$-axis corresponding to various discrete wavevector points between, say, $k_{BR}= \pi/\Lambda$ (edge of the Brillouin zone) and $0.5k_{BR}$? Or even more importantly in the context of optical telecommunications, e.g. DWDM components: What defect combinations for this lattice will provide localized photon states (i.e. high photon transmission) at desired optical frequencies, useful for multi-wavelength narrowband optical filters and tuneable semiconductor lasers?

The underlying physics behind the PBG effect and electromagnetic (EM) wave localization is the Bragg resonance between the wavevector $k$ of EM radiation and the lattice momentum $G$. In the simplest case of a periodic 1D lattice, composed of layers of two different materials with geometric thicknesses $H$ and $L$ and refractive indices $n_H$ and $n_L$ respectively, when the condition $\vec{k}_{BR}.\hat{G} = G_{BR}/2 = p/\Lambda$ is satisfied for a photon of wavevector $\vec{k}_{BR}$, a discontinuity (Bragg stop-band) forms which equates to a dip in transmission through the structure about an angular frequency $w_{BR}$ (the Bragg frequency) given by $w_{BR} = ck_{BR}$, where $c$ is the vacuum speed of light. We note the existence of the dominant Fourier component at an optical spatial frequency (momentum) given by $G_{BR} = 2p/\Lambda$, where $\Lambda = n_H H + n_L L$ is the periodicity in optical space. In our paper, real-space and reciprocal-space variables are all assumed the appropriate optical quantity[23] (i.e. optical space or path-length is given by $x = \int n(l) dl$ where $l$ is geometric real-space, and optical reciprocal-space is given by $G = 2p/x$), unless otherwise stated. It is widely known in digital signal processing (DSP) that the DFT of a real series produces a symmetric amplitude spectrum about the highest sampling frequency (known as the Nyquist frequency). This principle, taken together with the symmetry observed in the spectral responses of 1D quasiperiodic superlattices[11], suggests that the scattering sites of any real-space AL can be regarded as samples with a spatial frequency less than the highest spatial frequency $G_{BR}$ (the Nyquist frequency). Hence, the Fourier transform (FT) of an AL defined by a set of scattering sites $\{x_p\}$ will give a set of spatial frequency components $\{G_q\}$ that forms a symmetric spectrum about the highest spatial frequency $G_{BR}$, where the lattice reciprocal-space variable $G$ is the Fourier conjugate to the real-space variable $x$. Generalising the Bragg resonance condition between all of these spatial frequency components $\{G_q\}$ and the set of EM angular frequencies $\{w_q = cG_q/2\}$ gives us a qualitative determination of the spectral transmission (reflection) characteristic for EM wave propagation through the AL, indicated in Fig. 1. Clearly, this procedure results in transmission dips centred on the set of frequencies $\{w_q\}$, with symmetry about $w_{BR} = cG_{BR}/2$. Overall, this is an initial clue that the transmission characteristic of a photonic lattice is closely related to its spectral-distribution characteristics, i.e. the FT of its real-space structure (equivalent to the first Born approximation[24]). This FT link is the basis of our inverse design approach.

For mathematical convenience, the following DFT-based inverse analysis has been conducted in 1D only, however, we note that since the analysis is FT-based, it is easily extended to higher-dimension photonic lattices. We consider a binary relative permittivity lattice structure $e\{x_p\} = e_{ave} + \Delta e\{f_p\}$ of $N$ sites located at the set of positions $\{x_p\}$ in optical space. Each site is taken to be either $n_H$ or $n_L$, of respective geometric thicknesses $H$ and $L$. We use $e_{ave} = (n_H^2 + n_L^2)/2 = n_{ave}^2$ and $\Delta e = (n_H^2 - n_L^2)/2$. The set of parameters $\{f_p\} = \pm 1$ represents the binary lattice function (analogous to the Bravais lattice function for a periodic structure), where $f_p = +1$ equates to a high refractive index $n_H$, and $f_p = -1$ to the low refractive index $n_L$, so that defects inside the lattice can be controlled simply by manipulating the polarity $f_p$ of any individual site. Consider an arbitrary $h^{th}$ configuration of the photonic lattice denoted by $e_h\{x_p\}$. Taking the DFT of this $h^{th}$ configuration, yields its set of discrete spectral components $\bar{e}_h\{G_q\}$ given by:

$$\bar{e}_h\{G_q\} = \frac{\Delta e}{N} \sum_{p=1}^{N} f_p e^{-iG_q x_p} . \qquad (1)$$

Periodic boundary conditions define the set of discrete spatial frequencies $G_q$, given by

$$G_q = \frac{2q}{N} G_{BR}, \quad q = 1, 2, \ldots N/2 \qquad (2)$$

where $G_{BR}$ denotes the highest (Nyquist) spatial frequency $2p/\Lambda$. The maximum value of $q=N/2$ for the following analysis is sufficient due to the symmetric redundancy in the spectrum about the Nyquist frequency, discussed above. We can tailor the Fourier components distribution, given by equations 1 and 2 according to our need, and a subsequent inverse FT will generate a real-space lattice configuration. Various band properties, e.g. the strength of bandgaps or localized states, depend only on the amplitude of the associated Fourier component, with the phase characteristic therefore being a degree of freedom (i.e. arbitrary). This makes the inverse calculation analogous to the calculation of a computer-generated hologram (CGH)[25], with the formidable computational challenge of identifying those purely real CGH solutions from within the factorially-large search space. We use a non-deterministic simulated annealing (SA) optimization algorithm, to search this configuration space. By controlling the defects in the lattice its configuration is optimized when the cost function $E$ is minimized, where

$$E(\boldsymbol{e}_h) = \sum_{q=1}^{N/2} \left[ \left| \bar{\boldsymbol{e}}_{\text{target}}\{G_q\} \right| - \left| \bar{\boldsymbol{e}}_h\{G_q\} \right| \right]^2, \tag{3}$$

describes the 'error' between the lattice spectral response and the target spectral distribution $\bar{\boldsymbol{e}}_{\text{target}}\{G_q\}$. We note that multiple runs of the SA algorithm will tend to find different solutions, each of which is close to an overall global optimum in cost-space. From a practical point of view, the functionality of these solutions tends to be indistinguishable. As part of the design process, the appropriate value for $N$ is adopted to reflect the fabrication resolution and the overall physical size of the lattice, with $G$ existing in a quasi-continuous space as $N$ tends to infinity. Having generated an AL, we check that its transmission properties (i.e. band properties) are in agreement with the desired transmission characteristics. Conventional methods for this forward determination of the transmission characteristics of an AL use either transfer matrix (TM) methods, finite-difference time-domain (FDTD) approaches, or eigen-mode expansion (EME) techniques. However, we use the Fourier transform equation (4), which has been derived in Ref.[23] by solving the Ricatti equation for the scattering coefficient between a pair of forward- and backward-propagating coupled-modes, for quick and efficient (yet reasonably accurate) calculation of the transmission characteristics of our AL's.

$$\boldsymbol{t}(k) = \operatorname{sech}\left[ \left| \frac{1}{4n_{\text{ave}}^2} \int_{-\infty}^{\infty} \left( \frac{\partial \boldsymbol{e}(x)}{\partial x} \right) e^{-2ikDx} dx \right| \right]. \tag{4}$$

D is a modified Debye-Waller factor required to avoid the well-known phase accumulation error. Taking advantage of the efficient FFT algorithm, the full SA optimisation for a particular AL with $N$=112 took less than 1 minute using a Pentium IV processor with 2.8GHz clock frequency, and 512MB RAM. For larger $N$, the number of operations for the FFT scales as $N\log N$, and the overall SA optimisation scales accordingly. Higher dimensionality $d$ AL's simply scale as $dN^d \log N$.

Employing $N$=112 unit-cells, with the normalised DFT Nyquist frequency ($G_{BR}$) corresponding to $N/2 = 56$, we have 28 discrete points (indexed by the DFT variable $q$) between the edge of the Brillouin zone (corresponding to $k_{BR}$, $q$=56) and $0.5k_{BR}$, where $q$=28. To illustrate this, we present results for the first five points corresponding to $q$=56,55,54,53, and 52 respectively. We emphasize, though, that the lattice can be designed to exhibit a stop band 'on demand' at any position corresponding to $q$=1 through to $q$=56; however, space constraints allow us to show in Fig. 2 (i-v) only examples for the first five positions (the corresponding lattice configurations are presented in Table I). Figure 2(i) shows the spectrum for a conventional periodic lattice with a bandgap at the edge of the Brillouin zone (i.e. $q$=56). Transmission characteristics in Fig. 2 (ii-v) show the symmetric photonic stop gaps being shifted in incremental steps of about $0.02f_{BR}$,(i.e. $2f_{BR}/N$) away from the Brillouin zone edge, by using defects in well-defined locations in the photonic lattice. We also note the occurrence of resonant peaks between the two symmetric photonic stop bands. These resonant peaks form as a result of interference of different Bragg stop-bands, i.e., those Fourier components which make the lattice. Precise control of these peaks, important for DWDM components (e.g. when realised within the cavity of a semiconductor laser to enable tuning[26] ), thus requires tailoring of the relative strength of the different Bragg stop-bands. An important example is the so-called single-defect (0.5Λ) Fabry-Perot-type lattice[27] (used in a quarter-wave shifted DFB laser), corresponding to the $q$=55 AL. The high-frequency band-edge transmission peak of the lower bandgap (at $0.98f_{BR}$), and the low-frequency band-edge transmission peak of the upper bandgap (at $1.02f_{BR}$) overlap to form a single, very narrow resonant (localized) transmission peak at the Bragg frequency. In the example spectra of Fig.2(ii-v) we haven't tried controlling these peaks, as the intention is more to demonstrate the controlled tunability of Bragg stop-bands. Fig.2(vi), however, presents an AL which exhibits three high transmission (localized) states at $0.98 f_{BR}$, $f_{BR}$, and $1.02 f_{BR}$. We note that Fig.2(vi) can also be seen as an example of an AL exhibiting two Bragg stop-bands within the Brillouin zone. We also note that as the resulting complexity of the optical transmission characteristic is increased the individual Bragg stop-bands become less well defined. By introducing ever more defects, Fig. 2 shows a progression from a periodic to ever-more AL's, with the bandgap being shifted further away from the Brillouin zone edge. In so doing, the number of scatterers is reduced, so that the overall bandgap strength is reduced, but we note that we can straightforwardly increase the PBG bandgap strength again by simply increasing *N*, or the refractive index contrast Δ*e*.

In conclusion, we have presented a theory based on a Fourier space approach to achieve controllable field localization and photonic band properties using AL's. We believe this work will help PBG engineering find new applications, for example, with multiple defects introduced into a 3D photonic crystal fabricated using holographic lithography[29].

The authors thank P. B. Littlewood for helpful comments. S.C. thanks D. G. Hasko for discussions and support.

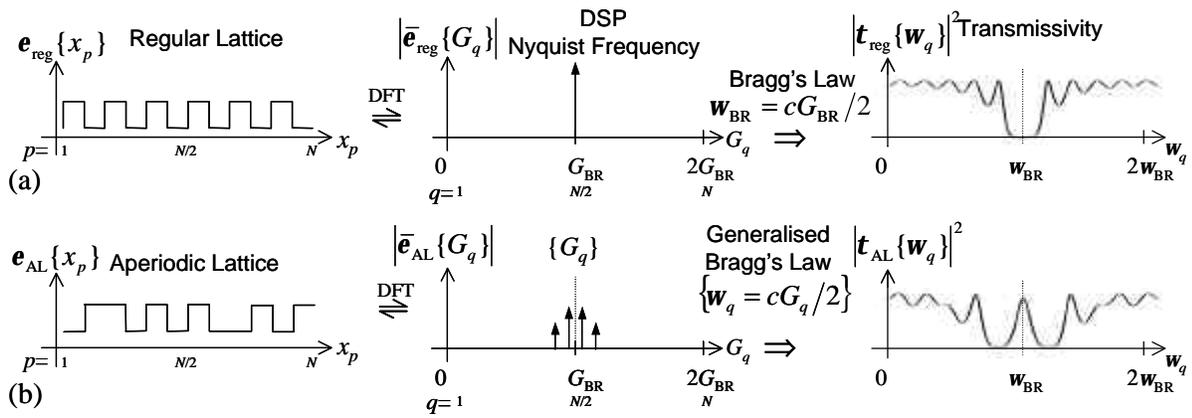

Fig.1: Qualitative understanding of how we can combine simple rules of DSP with Bragg's Law, in order to predict the position of the dips in the electromagnetic transmission spectra of any real AL.

Table 1. Optimised AL configurations

| Target Spatial Frequency index ($q$) | Detuning factor ($\Delta f/f_{BR}$) | High ($n_H$) and Low ($n_L$) RI combinations |
|---|---|---|
| 56 | 0 | $(HL)^{56}$ |
| 55 | 0.0179 | $(LH)^{28} (HL)^{28}$ |
| 54 | 0.0357 | $(HL)^{13} H^3 (HL)^{13} H (HL)^{13} H^3 (HL)^{13} L$ |
| 53 | 0.0535 | $(HL)^9 H (HL)^9 L (HL)^8 L^3 (HL)^8 H (HL)^9 H$ |
| 52 | 0.0714 | $(HL)^6 L (HL)^6 H^3 L^2 (HL)^5 L^3 (HL)^6 L (HL)^7 L$ $(HL)^6 H (HL)^6 L (HL)^7 L$ |

Tuning of photonic stop bands in incremental steps away from the conventional Bragg frequency $f_{BR}$, using $N=112$ unit-cells, with the normalised DFT Nyquist frequency ($G_{BR}$) corresponding to $N/2 = 56$.

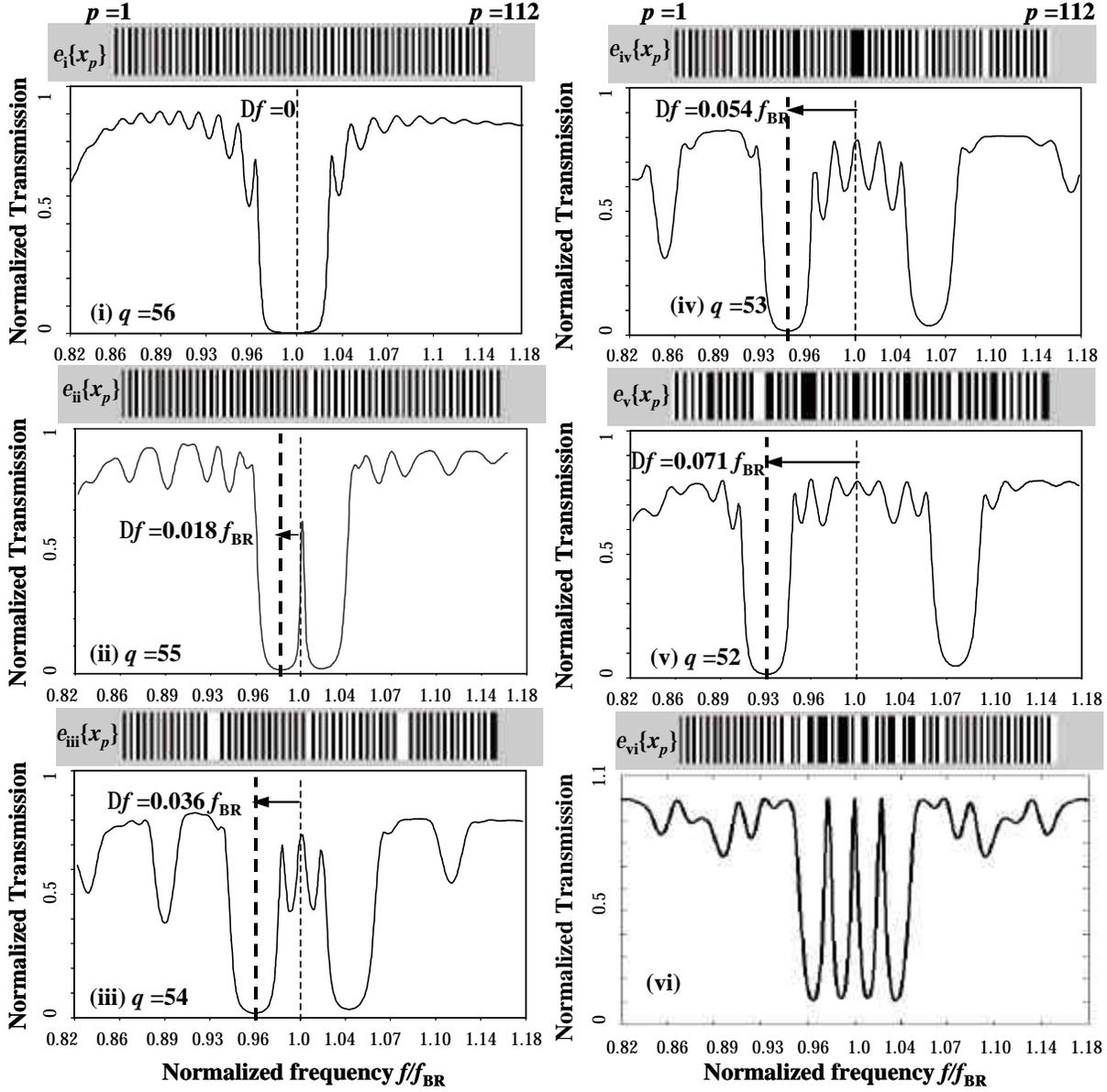

Fig.2. Optical transmission spectra (transmission $t$ versus normalized frequency $f/f_{BR}$), (i-v) calculated using EME[28], (vi) using equation 4. The aperiodic binary photonic lattices were designed in a rib-type single mode silicon waveguide by implementing the required changes in relative permittivity through binary modulation of the waveguide width. The effective refractive indices of the fundamental mode in the two widths of the waveguide were found to be $n_H=2.6$ and $n_L=2.42$ respectively. The optical path length for each waveguide-section was chosen to be a quarter Bragg-wavelength, with the Bragg wavelength $\lambda_{BR}=1550$nm, so that the two geometric lengths are $H\approx149$nm, and $L\approx160$nm respectively.